\documentclass{article}
\usepackage{graphicx}
\def\be{\begin{equation}}
\def\ee{\end{equation}}
\begin{document}

\title{On cosmic ray sources}
\author{G.Pizzella\\
Dipartimento di Fisica, Universit\`a di Roma ``Tor Vergata''\\
INFN Laboratori Nazionali di Frascati\\
{\rm e-mail: guido.pizzella@lnf.infn.it}}
\date{}
\maketitle

A recent paper \cite{pierre} on ultrahigh-energy cosmic rays (UHECR) data, with energies $\sim 10^{20}$ eV, obtained with the Pierre Auger Observatory during 3.7 years has shown that these cosmic rays have a non-isotropic spacial distribution. Quoting paper \cite{pierre}, \it the highest-energy particles originate from nearby extragalactic sources whose flux has not been substantially reduced by interaction with the cosmic background radiation. \rm  The Collaboration suggests that AGN or similar objects could be possible sources.

In this note I wish to call the attention that this result requires, for the cosmic ray acceleration, a mechanism different from that originally proposed by Fermi \cite{fermi}. As well known, in his original idea, Fermi considered an acceleration mechanism for cosmic rays based on the stochastic interaction of the charged particles with wandering interstellar magnetic field clouds. He obtained for the accelerated particles a power law energy spectrum, and he stressed the importance to have obtained a power law, consistently with the experimental data. In the Fermi model the power law exponent depends on the energy losses.

As well known, the experimental differential energy spectrum at high energy is of the type
\be
E^{-\gamma}
\ee
with $\gamma$ ranging $\sim 2.5-3$, for \bf all \rm charged particles (protons, electrons, heavier particles).

It is reasonable to think that if the Fermi theory applies also to the UHECR, one should expect an isotropic distribution for them, since the acceleration mechanism should operate everywhere in the intergalactic space.

At least another acceleration mechanism, acting locally, has been proposed in the past years \cite{pizze}. The basic idea was the following. The source is an astrophysical body surrounded by charged particles in a strong magnetic field (i.e. pulsars, AGN, magnetospheric objects) which we assume to have a dipolar nature. The particles are accelerated by electrical fields acting in the magnetosphere. The accelerated particles follow the magnetic lines of force towards the equator, with a motion that obeys the laws of the adiabatic invariance \cite{alfven}. At the equator the magnetic field is weakest along a given line of force. If the particle momentum does not satisfy anymore the requirements of the adiabatic invariance (because of the operating acceleration) then the particles leave the magnetosphere and go into space to become cosmic ray.

Simple calculations\footnote{
A charged particle  \cite{pizze} with electrical charge q and momentum p can be trapped
in a magnetic field with Larmor radius $R_L=\frac{p}{qB}$.
For a particle to stay trapped it is necessary that the Larmor radius be small
enough to satisfy the Alfven condition $\frac{\frac{dB}{dr}R_L}{B}\le \xi $.
The dimensionless quantity $\xi$ is estimated by means of plasma experiments
in space and on the Earth and is of the order of a few per cent. Combining the above equations we get $\frac{3pr^2}{qM}\le\xi$. The number of particles contained in a tube of force between the two
lines of force reaching the equator at distances r and r+dr is $dN\sim r^2 $. Combining the above equations  we get
\be
dN \sim p^{-2.5}dp
\ee
This is the number of particles with momentum between p and p + dp which
leave, with relativistic velocity, the corresponding tube of force. The maximum energy is given by
\be
E_{max}=\frac{\xi c q M}{3r_N^2}
\label{emax}
\ee
where $r_N$ is the radius of the source object.} give, in this model, a power law energy (momentum) spectrum. It is important to remark that here the exponent of the power law comes out in a very simple and natural way by geometrical considerations on the dipole magnetic field. This geometrical factor is independent on the species of the charged particles and it turns out to be exactly -2.5. A slight modification of this coefficient might occur during traveling in space.

We are aware that  we do not propose an explicit acceleration mechanism acting in the magnetosphere, but our considerations might help in the difficult task to fully understand the cosmic ray sources. 

As far as the maximum energy achievable, we find that a pulsar with a radius $r_N=10$ km and a magnetic field with dipole  moment $M=10^{21}~ Tesla~m^3$ gives $E_{max}\sim 5\cdot 10^{19}$ eV. We believe it might be possible to apply this model also to the case of an AGN.

 I thank Francesco Ronga and Rinaldo Santonico for useful discussions.

\end{document}